\documentclass[twocolumn,twocolappendix]{aastex631}

\begin{document}

\title{Properties of slow magneto-acoustic waves observed simultaneously using Hi-C 2.1 and AIA}

\author[0009-0009-5513-1001]{Suraj K. Tripathy}
\altaffiliation{Current Affiliation : Space Applications Centre, Indian Space Research Organisation, Ahmedabad 380015, India}
\email{surajkumartripathy703@gmail.com, suraj\_t@sac.isro.gov.in}
\affiliation{Indian Institute of Space Science and Technology, Thiruvananthapuram 695547, India}

\author[0000-0002-0735-4501]{S. Krishna Prasad}
\affiliation{Aryabhatta Research Institute of Observational Sciences, Nainital 263002, India}

\author{D. Banerjee}
\affiliation{Indian Institute of Space Science and Technology, Thiruvananthapuram 695547, India}
\affiliation{Center of Excellence in Space Sciences India, Indian Institute of Science Education and Research Kolkata, Mohanpur 741246, India}

\newcommand{\va}{46.0{\,}$\pm${\,}1.7{\,}km{\,}s$^{-1}$ }
\newcommand{\pa}{2.7{\,}$\pm${\,}0.2{\,}min }
\newcommand{\dpa}{4.0{\,}$\pm${\,}2.1{\,}Mm }
\newcommand{\daa}{3.4{\,}$\pm${\,}1.0{\,}Mm }
\newcommand{\lmba}{4.6{\,}$\pm${\,}0.7{\,}Mm }

\newcommand{\vh}{48.1{\,}$\pm${\,}0.6{\,}km{\,}s$^{-1}$ }
\newcommand{\ph}{2.8{\,}$\pm${\,}1.2{\,}min }
\newcommand{\dph}{4.1{\,}$\pm${\,}0.3{\,}Mm }
\newcommand{\dah}{3.7{\,}$\pm${\,}0.1{\,}Mm }
\newcommand{\lmbh}{4.8{\,}$\pm${\,}0.1{\,}Mm }



\begin{abstract}
Propagating slow magneto-acoustic waves are commonly observed in different coronal structures but are most prominent in active region fan loops. Their rapid damping with damping lengths of the order of a wavelength has been investigated in the past by several authors. Although different physical mechanisms have been proposed, significant discrepancies between the theory and observations remain. Recent high-resolution observations captured simultaneously by two different instruments reveal distinct damping lengths for slow magneto-acoustic waves although their passbands are similar. These results suggest a possible contribution of instrumental characteristics on the measurement of damping lengths. Here, we analyse the behavior of slow waves using a different pair of instruments in order to check the prevalence of such results. In particular, the cotemporal observations of active region NOAA AR12712 by the High-Resolution Coronal Imager (Hi-C 2.1) and the Atmospheric Imaging Assembly (AIA) onboard the Solar Dynamics Observatory (SDO) are utilised. The estimated oscillation periods of slow magneto-acoustic waves identified from these data are 2.7{\,}$\pm${\,}0.2{\,}min from SDO/AIA, and 2.8{\,}$\pm${\,}1.2{\,}min from Hi-C 2.1. The corresponding propagation speeds are found to be 46.0{\,}$\pm${\,}1.7{\,}km{\,}s$^{-1}$  and 48.1{\,}$\pm${\,}0.6{\,}km{\,}s$^{-1}$, respectively. Damping lengths were calculated by two different methods, the Phase Tracking Method (PTM) and the Amplitude Tracking Method (ATM). The obtained values from PTM are 4.0{\,}$\pm${\,}2.1{\,}Mm and 4.1{\,}$\pm${\,}0.3{\,}Mm while those from ATM are 3.4{\,}$\pm${\,}1.0{\,}Mm and 3.7{\,}$\pm${\,}0.1{\,}Mm, respectively, for the AIA and Hi-C data. Our results do not indicate any notable difference in damping lengths between these instruments.
\end{abstract}



\section{Introduction} \label{sec:intro}
Propagating intensity perturbations were first observed in the solar atmosphere along the coronal plumes by \citet{1998ApJ...501L.217D} using the 171{\,}{\AA} band of Extreme-ultraviolet Imaging Telescope (EIT) onboard Solar and Heliospheric Observatory (SOHO). These perturbations were found to travel at speeds approximately equal to the local sound speed. Hence, they were identified as slow magnetoacoustic waves. 
Later, these waves were observed in active region fan loops by various instruments such as SOHO/EIT \citep{1999ESASP.446..173B}, Transition Region and Coronal Explorer (TRACE) \citep{2000A&A...355L..23D}, the Atmospheric Imaging Assembly (AIA) onboard the Solar Dynamics Observatory (SDO) \citep{2012SoPh..281...67K} etc. 

Previously, different authors have calculated the wave parameters such as oscillation period, propagation speed, amplitude, etc., for the slow magneto-acoustic waves observed in coronal loops. \citet{2009SSRv..149...65D} presented a statistical overview of such parameters calculated primarily using the data from TRACE 171{\,}{\AA}. The range of values for the oscillation period, propagation speed, and amplitude (as a percentage of the background) of these waves were found to be 145--550 s, 45--205 km{\,}s$^{-1}$ and 0.7--14.6\%, respectively. 
Moreover, \cite{2002A&A...387L..13D} found a relation between the oscillation period and the location of the footpoints of the associated coronal loops. Their results reveal that the loops rooted in sunspots display smaller periods i.e., around 3 minutes while those rooted in plages exhibit slightly larger periods around 5 minutes. Subsequently, similar association has been observed by various authors.

Interestingly, the amplitude of the perturbations was observed to be very rapidly decaying resulting in the disappearance of waves within a short distance ranging from 2.9 to 23.2 Mm \citep{2002SoPh..209...61D}. This characteristic distance is known in the literature by various names such as decay length, e-folding length, damping length, etc. In this work, the authors have decided to refer to this as damping length. The rapid decay in wave amplitude has been attributed to various physical mechanisms including thermal conduction, compressive viscosity, and optically thin radiation in addition to the geometric effects like magnetic field divergence. Among these, the dominant contributor was found to be the thermal conduction \citep{2003A&A...408..755D,2004A&A...415..705D}.

\begin{figure*}
\includegraphics[width=\textwidth]{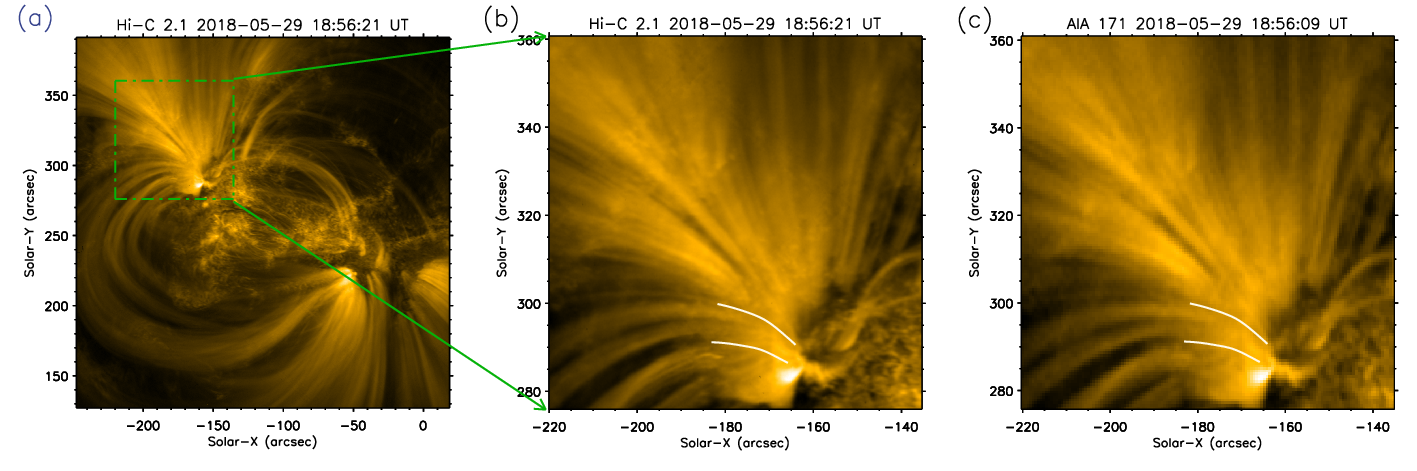}
\caption{(a) Full field of view of Hi-C 2.1 observing the active region AR12712 in the 172{\,}{\AA} band. The box marked by green dashed lines outlines the subfield region presented in the other two panels. (b) and (c) A zoomed-in view of the loop structures in the vicinity of the target structure as observed from Hi-C and AIA, respectively. The white curves marked in these panels represent the boundaries of the selected loop where propagating slow magneto-acoustic waves are found.}
\label{fig:1st}
\end{figure*}

It may be noted that a discrepancy over two orders of magnitude exists between the observed damping length values and those obtained from the theory of thermal conduction \citep{2019FrASS...6...57S}. The mismatch is significant even after incorporating damping due to additional effects. A part of this is generally attributed to projection effects (as the observed loops do not necessarily align with the plane perpendicular to our line-of-sight, any deviation from it would result in all length measurements including the damping lengths being shortened because of projection). 

The actual deprojected damping length of slow waves was calculated for the first time by \citet{2011ApJ...734...81M} using stereoscopic observations. For the case that these authors studied, the damping length value was obtained to be about \text{$20^{+4}_{-3}$} Mm. Further investigation revealed that the dominant contributor to the amplitude decay in this particular case was magnetic field divergence, instead of thermal conduction. The expected damping length considering the magnetic field divergence was estimated to be about 55 Mm, which is still a factor 2 to 3 higher than the observed value. So, there remains a clear gap in our understanding of the damping of slow waves.

Furthermore, adding to the complexity, recently, \citet{2024MNRAS.527.5302M} have reported significantly different damping lengths for slow waves detected within the same loop when observed using two different instruments, namely, the SDO/AIA and the Extreme Ultraviolet Imager (EUI) onboard the Solar Orbiter (SolO). At the time of this observation, the viewing angles of these two telescopes were only slightly different with a separation of about 19$^{\circ}$ which could not explain the discrepancy. This is a quite surprising result, and the physical reason behind it still remains unknown. But, besides probing any scientific explanation, the prevalence of such observations must also be checked. Therefore, in this work, we calculate and compare the damping lengths of slow waves observed from another pair of instruments. We present the details on the observations and describe the analysis techniques used along with a discussion of the results found in the following sections.

\section{Observations} \label{sec:observ}
The data used in this study are comprised of imaging sequences of active region AR 12712 captured using two different telescopes, namely, the High-Resolution Coronal Imager \citep[Hi-C;][]{2019SoPh..294..174R} and the Atmospheric Imaging Assembly \citep[AIA;][]{2012SoPh..275...17L}. 

The Hi-C telescope was launched on a sounding rocket on 29th May, 2018. A series of similar rocket launch experiments have been performed in the past under the same name, so the data from the particular iteration we use here are labelled as Hi-C 2.1. As a part of this experiment, during the flight, Hi-C captured a series of images over a small field of view (4.4$^{\prime} \times$4.4$^{\prime}$) encompassing AR 12712 for about 335{\,}s starting from 18:56:21 $\--$ 19:01:56 UT. For these observations, an Extreme UltraViolet (EUV) imaging filter with its passband peaking at 172{\AA} is utilised.  
 The time cadence is at 4.35{\,}s and the plate scale is 0.129$^{\prime\prime}$. 
The publicly available, level 1.5 data of this experiment were downloaded from the MSFC/NASA site\footnote{\url{https://msfc.virtualsolar.org/Hi-C2.1/}}. These data were already calibrated for basic-instrument corrections and are considered science ready. However, the last few frames display a sharp rise in intensity potentially suggesting a bit of overcompensation for atmospheric absorption in these images. We exclude these images shortening the duration to 304 s. 
Addtionally, as mentioned in the Hi-C user guide about half of the images are smeared due to substantial jitter from the rocket. To minimise the effect of these frames in the data co-alignment, we rebin the data averaging every two successive frames resulting in a final cadence of 8.7 s.

The AIA telescope on board the Solar Dynamics Observatory \citep[SDO;][]{2012SoPh..275....3P} observes the sun in several Extreme Ultra-Violet (EUV) channels out of which the data corresponding to a channel similar to that of Hi-C, centred at 171{\,}\AA\ are used. Since the AIA observes the full disk of the sun at all times, a subfield region coinciding with the Hi-C field of view is extracted. Keeping in mind the short duration of the Hi-C data, an image sequence of 30 minutes duration starting from 18:56:09 UT, is selected. All the AIA data were downloaded and calibrated to the science-ready level 1.5 quality using a robust pipeline developed by Rob Rutten \footnote{\url{https://robrutten.nl/rridl/sdolib/dircontent.html}}. 
 These data have constant cadence of 12 s and a plate scale of 0.6$\arcsec$.

In order to study the oscillations and compare their properties across the two datasets, it is important to select the same loop region in both the datasets. A proper coalignment between them is therefore a prerequisite, to ensure which we implement the following steps. First the AIA data were internally coaligned to match the features in the first frame. Second, the Hi-C data were derotated by 0.985$^\circ$ to correct for roll angle relative to AIA. A desired subfield region encompassing the location of the detected oscillations is then extracted from both the datasets and subsequently, the AIA data are upscaled to match the Hi-C pixel scale. In the final step, each image from Hi-C data is cross-correlated with the closest AIA image and the obtained shifts are applied to the Hi-C data. A similar coalignment technique had been implemented earlier by \citet{2020ApJ...896...51W}. Note that the upscaling of AIA data is done only for coalignment purposes and rest of the analysis is done keeping the data at original resolution. The selected subfield region and the appearance of loop structures within both the datasets are shown in Fig. \ref{fig:1st}.

\begin{figure*}
\includegraphics[width=\textwidth]{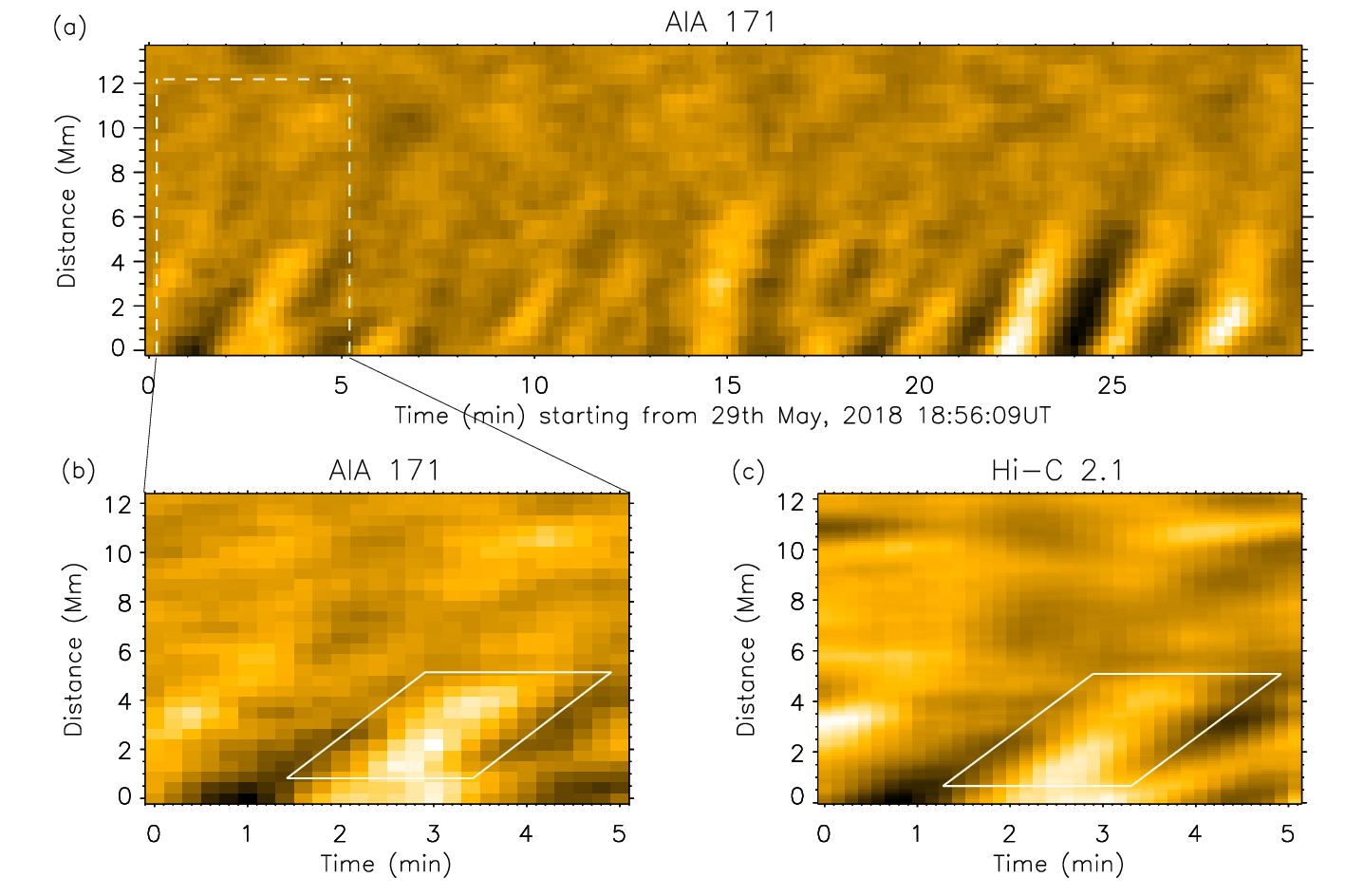}
\caption{Time-Distance map constructed from the loop marked in Figure \ref{fig:1st} for AIA (a) in its full duration, (b) in the duration overlapping Hi-C, (c) for Hi-C. The white dashed lines in panel a) outline the region overlapping Hi-C as shown in the bottom two panels. The white parallelogram in panels b) and c) marks the region used for computing the propagation speed.}
\label{fig:td}
\end{figure*}

\section{Analysis \& Results}
We search for slow magneto-acoustic waves in the data by inspecting the intensity evolution along individual loop structures. We perform this by constructing a time-distance map along each loop structure and see if there are any oscillation patterns present in them. For the construction of time-distance maps, first, the boundaries of a loop are chosen. Then, the intensities across the loop are averaged to get a 1D intensity profile. Similar profiles built from this loop region from all the images across the full time sequence are then stacked together to obtain a time-distance map. The amplitude of the oscillations that we are trying to find is expected to be small, typically about 5\% of the background, hence, a background subtraction would be necessary to enhance their appearance. For the AIA data, this background was constructed by smoothing the time series over a window of 5 minutes. Because the Hi-C data is only five minutes long, an average over the full duration is considered for the estimation of background. The detrended time-distance maps are then constructed by subtracting the respective backgrounds at each spatial location along the loop. Finally, to remove high frequency noise, the detrended time-distance maps are smoothed over a window of width around 30\,s.

The presence of slanted ridges of alternating brightness in the time-distance maps indicates outward propagating waves. Ideally, the higher spatial and temporal resolution Hi-C data are considered better suitable for the initial search but because of their shorter duration, we perform the analysis using the AIA data first and if any oscillatons are visible in the first five minutes, we look for corresponding patterns in the Hi-C data. Although we inspected several loops, the three-minute waves were found to be prominent in both AIA and Hi-C data in only one loop structure which is marked in Fig. \ref{fig:1st}. The corresponding detrended time-distance maps obtained from both the datasets are shown in Fig. \ref{fig:td}. Similar oscillatory patterns can be seen in the both the maps. The oscillations are visible only upto a distance of 7 Mm or so beyond which their presence could not be clearly discerned. The extraction of oscillation parameters like time period, propagation speed, and damping length, from these time-distance maps, is described in the following subsections and the obtained values are summarized in Table \ref{tab:all}.

\begin{figure*}
\includegraphics[width=\textwidth]{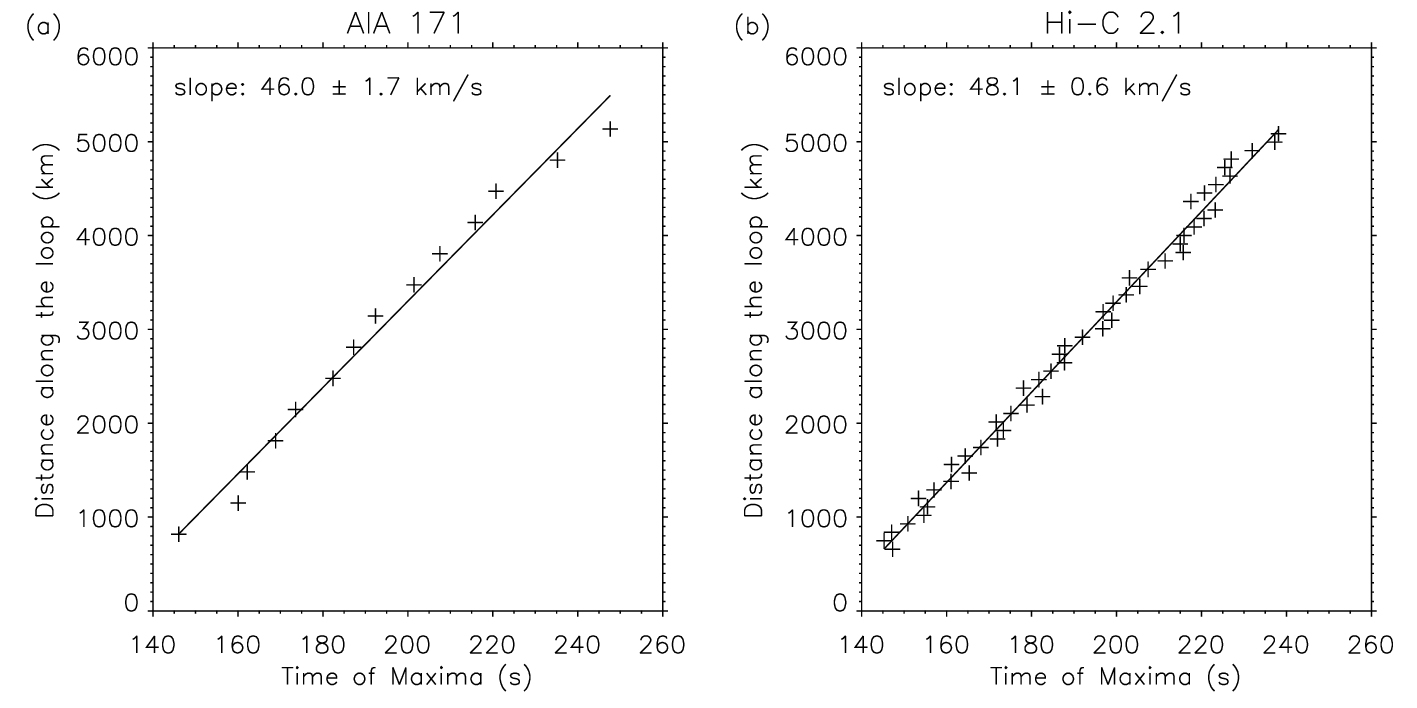}
\caption{Propagation speed estimation. The + symbols denote the temporal locations of the maxima as a function of distance along the bright ridge isolated by a white parallelogram in Fig. \ref{fig:td}. The solid line corresponds to a linear fit to the data. The slope obtained from the fit, representing the propagation speed, is listed in the plot. Panels (a) and (b) depict the data for AIA and Hi-C respectively.}
\label{fig:velo}
\end{figure*}

\subsection{Time period}\label{tp}
For the estimation of oscillation period, we subject the detrended intensity time series at each spatial location to Fourier transform using a Fast Fourier Transform (FFT) code. Following this, the power spectra at all spatial locations within a distance of $\approx$7 Mm (upto where the oscillations are visible) are averaged to obtain a mean power spectrum. From this spatially averaged spectrum, the dominant frequency (and hence the dominant period) is identified by fitting the strongest peak in power to a Gaussian. The standard deviation of the fitted Gaussian is then taken as the corresponding error. The obtained period values are \pa, and \ph, respectively, for the AIA and Hi-C data. Note, the larger error in the Hi-C value is because of the lower resolution in its power spectra, a consequence of the shorter duration of the data.

\begin{figure*}
\includegraphics[width=\textwidth]{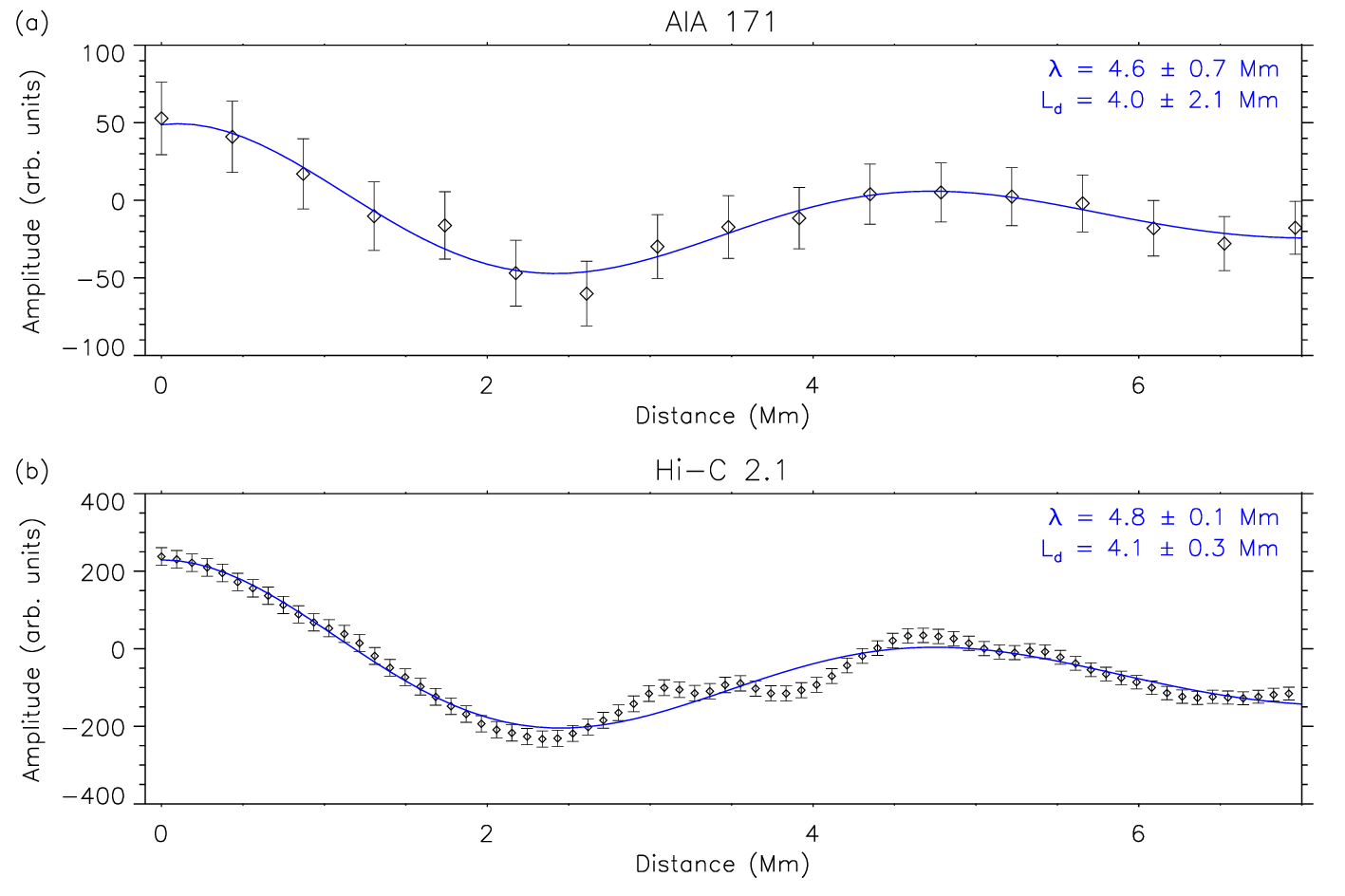}
\caption{Extraction of wave parameters using the phase tracking method. The diamond symbols denote the intensities at a fixed temporal location marked by a black dashed line in Fig. \ref{fig:td}. The error bars denote the respective uncertainties. The best-fit damping sinusoid function to these data is shown as a blue solid line. The corresponding wavelength ($\lambda$) and damping length ($L_d$) are listed in the plot. The top and bottom panels depict the data from AIA and Hi-C, respectively.}
\label{fig:ptm}
\end{figure*}

\begin{figure*}
\includegraphics[width=\textwidth]{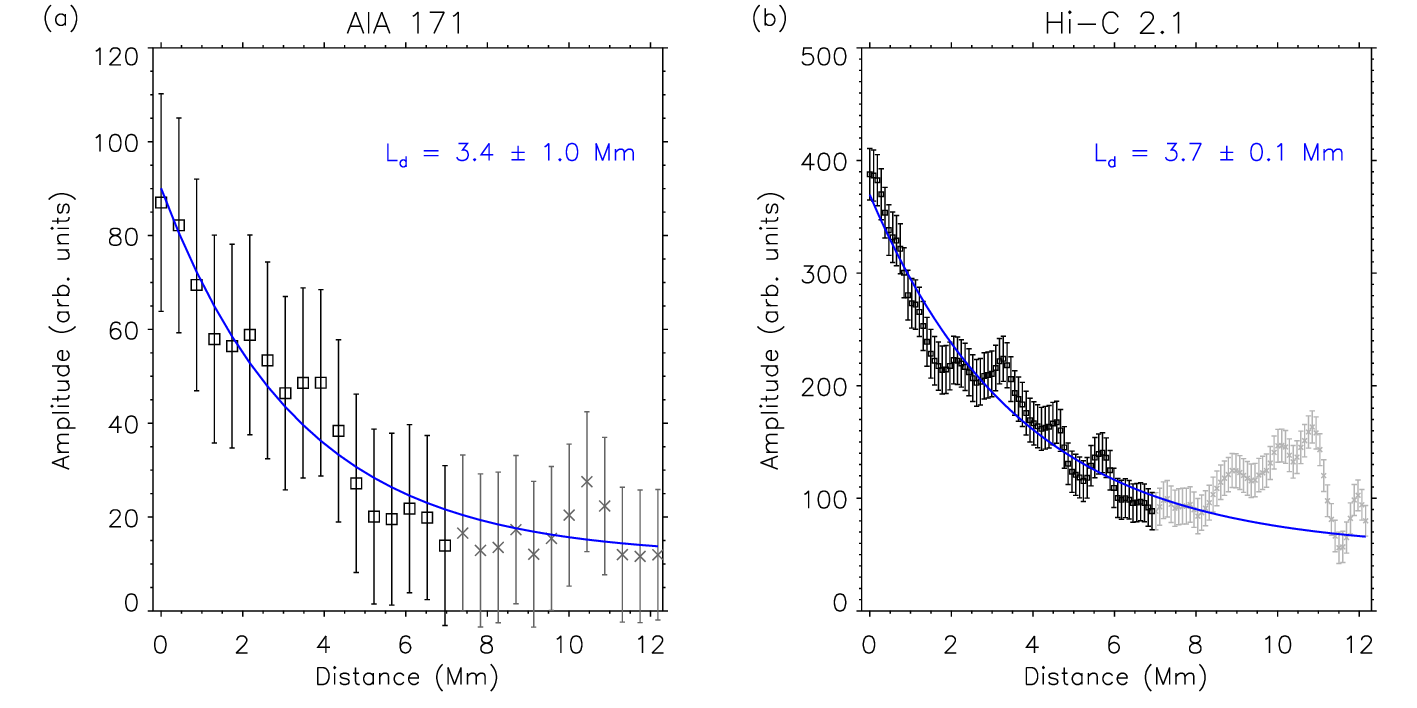}
\caption{Extraction of damping length using amplitude tracking method. The square and 'x' symbols mark the amplitude of the oscillation as a function of distance along the loop. The error bars denote the respective uncertainties. The blue curve represents the best-fit decaying exponential function to the data. The amplitude at the last spatial location is used as the minimum amplitude for the fit. The amplitude data shown with 'x' symbols in grey are excluded from the fit as the oscillations are not visible in this region. The obtained damping length values are listed in the plot. Panels (a) and (b) show the data for AIA and Hi-C, respectively.}
\label{fig:atm}
\end{figure*}

\subsection{Propagation speed}
The slope of the slanted ridges in the time-distance maps gives the propagation speed. In order to estimate this, we consider the central bright ridge identified by a parallelogram in Fig. \ref{fig:td}. Within this region, the intensity profile across the ridge is extracted from each spatial location, and the corresponding temporal location, where the local maximum is present, is noted. The top and bottom boundaries along the ridge are restricted to ensure proper identification of the local maxima. The obtained temporal location (in seconds) from the beginning of the time series is plotted as a function of distance (in km) along the loop in Fig. \ref{fig:velo} for both datasets. The best linear fit to these data obtained from chi-square minimisation is shown as a solid line in this figure. The calculated slope value from the fitted line gives a propagation speed of \va for AIA and \vh for Hi-C data. The errors are estimated from the uncertainty in the fitted slope. These values imply a close agreement between the propagation speeds from both datasets. In addition, the obtained speeds are much lower than the local sound speed, which is about 132 km{\,}s$^{-1}$ (assuming a peak response temperature of 0.8 MK). This discrepancy is because of the projected distances used in the calculation of the speed. Furthermore, the magnitude of the difference in the values indicates that the observed loop is only slightly inclined with respect to the line of sight with an inclination angle of 21$^{\circ}$. Such small inclination angles are expected in the vicinity of a sunspot where the target loop structures are rooted.

\subsection{Damping length}
The slow magneto-acoustic waves are known to get rapidly damped. In the present case, the oscillatory signatures seem to disappear after a distance of 7 Mm or so from the loop foot point. In order to quantify how quickly they are getting damped, we estimate the damping length, a characteristic length over which the oscillation amplitude decreases by a factor of `e' ($\approx$2.7), using two different methods similar to that explained in \citet{2019FrASS...6...57S}.
\subsubsection{Phase Tracking Method (PTM)}
In this method, we follow the evolution of wave amplitude by tracking the phase of the wave. This is done by extracting spatial intensity profiles at a fixed time frame from the time-distance maps. The intensity profiles extracted from both the datasets from the same time step are shown in Fig. \ref{fig:ptm} using diamond symbols. The corresponding uncertainties in the data are estimated by assuming a data noise dominated by photon noise and readout noise as described in Appendix \ref{sec:appx}. The intensities only upto 7 Mm are considered as the oscillations are barely discernible at larger distances. These data are then fitted to an exponentially decaying sinusoid of the form, 
\begin{equation}\label{eq:ptm}
    I(x)= A_{0}\exp{\left( \frac{-x}{L_{d}} \right)} \sin\left(\frac{2\pi x}{\lambda}+\phi \right)+B_{0}+B_{1}x
\end{equation}
where, $x$ is the distance along the loop, $A_0$ is the maximum amplitude, $L_d$ is damping length, $\lambda$ is wavelength, $\phi$ is the inital phase, and $B_0$ and $B_1$ are appropriate constants. The best fits obtained by chi-square minimisation are shown by solid curves in Fig. \ref{fig:ptm}. The corresponding wavelength values are \lmba and \lmbh and the damping length values are \dpa and \dph, respectively, for the AIA and Hi-C data. These values indicate a good agreement (within uncertainty) between both datasets.

\subsubsection{Amplitude Tracking Method (ATM)}
\label{subsec:atm}
In this method, instead of extracting the intensity profile from a particular time step, we directly estimate the average amplitude at each spatial location and see how it evolves as a function of distance. This method takes advantage of the fact that for a purely sinusoidal signal, the standard deviation $\sigma$ and the amplitude $A$ are related by, $A=\sqrt{2} \sigma$. A potential drawback of this method is that the presence of any localised brightenings in the time-distance maps would contaminate the amplitudes. This effect would be important especially when the time series is short. Indeed  a localised brightening may be seen in the time-distance map of Hi-C (see Fig. \ref{fig:td}c) at the beginning the time series between a distance of 2-4 Mm. In order to avoid the influence of this, we exclude the first 45 s of the time series from the amplitude calculations. Although the brightening is not as prominent in the AIA data, to keep the calculations consistent, we apply the same for AIA data as well. The standard deviation of intensities within rest of the time series is computed at each spatial location and the amplitude computed from this, as described above, is plotted as a function of distance in Fig. \ref{fig:atm} for both datasets. The uncertainty on the amplitude is obtained from the median value of the uncertainty on intensities (see Appendix \ref{sec:appx}) at each spatial location. These data are then fitted to an exponentially damped signal of the form,
\begin{equation}\label{eq:atm}
    A(x)= A_{0}\exp{\left(\frac{-x}{L_{d}}\right)}+C
\end{equation}
where, $x$ is the distance along the loop, $L_d$ is the damping length, $A_0$ is the maximum amplitude, and $C$ is an appropriate constant. The value of C, for each case, is fixed as the amplitude at the last spatial position i.e., at a distance of $\approx$ 12 Mm. The best fits obtained from the chi-square minimisation are shown by solid blue curves in Fig. \ref{fig:atm}. Because the oscillations are only visible up 7 Mm or so, the data beyond this height (shown in grey in the plot) are not used in the fit. The final damping length values obtained are \daa and \dah for the AIA and Hi-C data, respectively. These values are in good agreement with each other and also match closely with those obtained from the phase tracking method.

\section{Discussion \& Conclusions}
In this study, we analyse and compare the properties of slow magneto-acoustic waves observed simultaneously using the space-based AIA data and the rocket flight Hi-C 2.1 data. Considering the short duration ($\approx$ 5 min) of the Hi-C data, our main target was to detect  3-minute oscillations that are typically found in active region loops rooted in sunspots. Although several such loops were analysed, we could only find one example where the oscillations are clearly visible in both Hi-C and AIA data. To our knowledge, this is the first time the detection of slow waves is reported from the Hi-C data. 

The time-distance maps constructed from both datasets show outward propagating compressive waves exhibiting similar behaviour (see Fig. \ref{fig:td}). The oscillation period was found to be \pa from the AIA data and \ph from the Hi-C data. The corresponding propagation speeds are obtained as \va and \vh. The similarity between these values provides additional confirmation that we are looking at the same oscillations from both instruments.

The oscillations disappear after a short distance ($\approx$ 7 Mm) from the loop foot point. To quantise this damping behaviour, we compute damping lengths using two different methods, namely, the phase tracking method and the amplitude tracking method. The damping lengths obtained from the phase tracking method are \dpa for AIA and \dph for Hi-C, while those from the amplitude tracking method are \daa for AIA and \dah for Hi-C. All these values are summarized in Tab. \ref{tab:all}. Although each of the employed methods has its own limitations (for instance, the phase tracking method requires the manual selection of a temporal location while the amplitude tracking method assumes a perfect sinusoidal signal), we do not find any significant discrepancy between the damping length values obtained from both datasets. Furthermore, the time-distance maps also highlight the visibility of oscillations up to similar length scales in both datasets.

Similar attempts to study the properties of slow waves observed using different instruments were made in the past \citep{2001A&A...370..591R, 2003A&A...404L...1K, 2003A&A...404L..37M} albeit using different passbands. Although these authors do not explicitly calculate the damping lengths, the time-distance maps presented in those articles do not show any significant difference in the length scales over which the oscillations are visible. Later, \citet{2009ApJ...697.1674M} and \citet{2011ApJ...734...81M} have studied the slow waves observed using the identical passbands but from different vantage points. These observations were carried out using the Extreme UltraViolet Imager (EUVI) onboard the Solar Terrestrial Relations Observatory (STEREO) A and B spacecrafts. The authors estimate a damping length of 20$^{+4}_{-3}$ Mm for the STEREO-A data and 27$^{+9}_{-7}$ Mm for the STEREO-B data. Again, these values are not notably different considering the large uncertainties. 

However, in contradiction to all these results (including ours), a more recent study by \citet{2024MNRAS.527.5302M} finds a noticeable difference in the damping length of slow waves observed using SolO/EUI and SDO/AIA. The authors obtain damping lengths of 12.8$^{+1.1}_{-1.7}$ Mm and 6.9$^{+1.3}_{-0.8}$ Mm, respectively, for the EUI and AIA data. As stated earlier, the angle of separation between these instruments at the time of observation is about 19$^\circ$. The authors demonstrate that even after accounting for this viewing angle separation, the observed discrepancy in damping length could not be explained. Another factor to note is that there is a marginal difference in the temperature response of the passbands used by EUI and AIA \citep{2021A&A...656L...7C}. In comparison, in the present case, the viewing angle remains the same for the Hi-C and AIA, but there is still a minor difference in the temperature response (although the peaks coincide) of the respective passbands \citep{2019SoPh..294..174R}. Therefore, it is not trivial to conclude if a small difference in the temperature response of the passbands can result in a large difference in observed damping lengths. Overall, with these observations alone, it is difficult to isolate the cause for obtaining instrument dependent damping length for slow waves. We believe more such joint observations in the future can help in resolving this puzzle.

\begin{table}[]
    \movetableright=-1.25cm
    \centering
    \caption{Summary of computed slow wave parameters}
    \begin{tabular}{ccc}
    \hline
         Parameters & SDO/AIA &  Hi-C 2.1 \\ \hline \hline
         Oscillation period & \pa  & \ph \\ 
         Propagation speed & \va & \vh \\
         Damping length - PTM & \dpa & \dph \\ 
         Damping length - ATM & \daa & \dah \\
         \hline
    \end{tabular}
    \label{tab:all}
\end{table}

\section*{Acknowledgments}
We acknowledge the High-resolution Coronal Imager (Hi-C 2.1) instrument team for making the second re-flight data available under NASA proposal 17-HTIDS17\_2-003. MSFC/NASA led the mission with partners including the Smithsonian Astrophysical Observatory, the University of Central Lancashire, and the Lockheed Martin Solar and Astrophysics Laboratory. Hi-C 2.1 was launched out of the White Sands Missile Range on 2018 May 29. The AIA data used are the courtesy of NASA/SDO and the AIA science team. SKP is grateful to SERB/ANRF for a startup research grant (No. SRG/2023/002623).

\appendix

\section{Error estimation}
\label{sec:appx}
We follow the procedure described below for the estimation of uncertainty on intensities in time-distance maps. Assuming the uncertainty in intensity is dominated by the photon noise and the readout noise, we can write
\begin{equation}\label{eq:tot_noise}
    \sigma_{noise}^2(I)= {\sigma_{readout}^2+ \sigma_{photon}^2(I)}
\end{equation}
where, $\sigma_{noise}$ is the total noise equivalent to the final uncertainty on intensity, $\sigma_{readout}$ is the readout noise and $\sigma_{photon}$ is the photon noise. This assumption is valid as long as the signal-to-noise ratio in our region of interest is good. For the low signal locations, noise due to additional sources shall be added. Since our target region in this work generally displays high intensities, ignoring all the other noise terms seems reasonable.

The photon noise varies with the observed intensity. Therefore, for the computation of it, we use
\begin{equation}\label{eq:pho_noise}
    \sigma_{photon}^2(I)= G*I
\end{equation}
where, $I$ is the intensity in DN and $G$ is the detector gain in DN/photon. Note the intensity within time-distance maps corresponds to an average value across the loop so an appropriate error propagation is applied. The readout noise, on the other hand, is fixed and the value of it is usually provided by the instrument teams. The detector gain is also a fixed value.

The readout noise and gain for the SDO/AIA 171 \AA\ channel are 1.15 DN and 1.168 DN/photon, respectively \citep{2012SoPh..275...41B}. The readout noise for the Hi-C 2.1 data is 4.0 DN \citep{2019SoPh..294..174R}. However, the gain for Hi-C data is not provided. We consider this as unity here.








\end{document}